\documentclass[a4paper,12pt]{article}
\pdfoutput=1
\usepackage{epsfig}
\usepackage{amsmath}
\usepackage{amssymb}
\usepackage{amsfonts}
\usepackage{bbm}
\usepackage{fleqn}
\usepackage[footnotesize]{caption}
\usepackage{graphicx}
\usepackage{mathrsfs}
\usepackage[center,footnotesize,hang]{subfigure}

\def\be{\begin{equation}}
\def\ee{\end{equation}}
\def\beq{\begin{equation}}
\def\eeq{\end{equation}}
\def\bea{\begin{eqnarray}}
\def\eea{\end{eqnarray}}

\def\<{\left\langle}
\def\>{\right\rangle}

\def\be{\begin{equation}}
\def\ee{\end{equation}}
\def\beq{\begin{equation}}
\def\eeq{\end{equation}}
\def\bea{\begin{eqnarray}}
\def\eea{\end{eqnarray}}

\newcommand{\newc}{\newcommand}
\newc{\ol}{\overline}
\newc{\wt}{\widetilde}
\newc{\bs}{\boldsymbol}
\newc{\ma}{\mathcal}
\newc{\vl}{\langle}
\newc{\vr}{\rangle}
\def\vev#1{\langle #1 \rangle}
\newc{\sg}{S}
\newc{\ug}{U}
\newc{\tg}{T}

\allowdisplaybreaks[2]
\begin{document}
\bibliographystyle{OurBibTeX}

\title{\hfill ~\\[-30mm]
                  \textbf{Minimal see-saw model predicting best fit lepton mixing angles}        }
\date{}
\author{\\[-5mm]
Stephen F. King\footnote{E-mail: {\tt king@soton.ac.uk}}\\ \\
  \emph{\small School of Physics and Astronomy, University of Southampton,}\\
  \emph{\small Southampton, SO17 1BJ, United Kingdom}\\[4mm]}

\maketitle

\begin{abstract}
\noindent
{
We discuss a minimal predictive see-saw model in which the right-handed neutrino mainly responsible for the atmospheric neutrino mass has couplings to $(\nu_e, \nu_{\mu}, \nu_{\tau})$ proportional to $(0,1,1)$ and the 
right-handed neutrino mainly responsible for the solar neutrino mass has couplings to $(\nu_e, \nu_{\mu}, \nu_{\tau})$ proportional to $(1,4,2)$, with a
relative phase $\eta = -2\pi/5$.
We show how these patterns of couplings could arise from an $A_4$ family symmetry
model of leptons, together with $Z_3$ and $Z_5$ symmetries which fix  $\eta = -2\pi/5$ 
up to a discrete phase choice. The PMNS matrix is then completely determined by one remaining parameter 
which is used to fix the neutrino mass ratio
$m_2/m_3$. The model 
predicts the lepton mixing angles 
$\theta_{12}\approx 34^{\circ}, \theta_{23}\approx 41^{\circ}, \theta_{13}\approx 9.5^{\circ}$,
which exactly coincide with the current best fit values for a normal neutrino mass hierarchy,
together with the distinctive prediction for the CP violating oscillation phase $\delta \approx 106^\circ$. } 
 \end{abstract}
\thispagestyle{empty}
\vfill
\newpage
\setcounter{page}{1}

\section{Introduction}

Daya Bay  \cite{An:2012eh} and RENO \cite{Ahn:2012nd} have measured a non-zero reactor angle 
$\theta_{13}\approx 0.15$ which excludes Tri-bimaximal (TB) mixing \cite{Harrison:2002er}.
Recent global fits also hint at deviations of the atmospheric and solar angles from their TB values
(for a recent review see e.g. \cite{King:2013eh}).
Such deviations may be expressed in terms of the deviation parameters ($s$, $a$ and
$r$) from TB mixing~\cite{King:2007pr} (for a related parametrisation see \cite{Pakvasa:2007zj}): 
\be
\label{rsadef}
\sin \theta_{12}=\frac{1}{\sqrt{3}}(1+s),\ \ \ \ 
\sin\theta_{23}=\frac{1}{\sqrt{2}}(1+a), \ \ \ \ 
\sin \theta_{13}=\frac{r}{\sqrt{2}}.
\ee
With zero solar and atmospheric deviations from TB mixing, $s=a=0$,
and Cabibbo-like reactor mixing described by $r=\lambda $, 
with $\lambda = 0.225$ being the Wolfenstein parameter, one is led to
Tri-bimaximal-Cabibbo (TBC) mixing~\cite{King:2012vj}.
However, as mentioned above, current global fits prefer non-zero solar and atmospheric TB deviation parameters,
\beq
s=-\lambda^2/2, \ \ \ \ a=-\lambda /3, \ \ \ \ r=\lambda ,
\label{ansatz}
\eeq
corresponding to the angles,
\beq
\theta_{12}=34.2^{\circ}, \ \ \ \ \theta_{23}=40.8^{\circ},\ \ \ \   \theta_{13}=9.15^{\circ}.
\label{bestfitangles}
\eeq
These angles are close to the best fit values for all three global fits in the case of a normal neutrino mass ordering
\cite{King:2013eh}.
Assuming a normal neutrino mass {\em hierarchy} with $m_1=0$, one is led to \cite{King:2013iva},
\beq
\frac{m_2}{m_3}=\frac{3}{4} \lambda,
\eeq
corresponding to $m_2/m_3 \approx 0.17$, close to the best fit value \cite{King:2013eh}.
The deviation parameters in Eq.\ref{ansatz} 
have the feature that the atmospheric mixing angle is in the first octant and the solar mixing angle
is somewhat less than its tri-maximal value, in agreement with 
the latest global fits for the case of a normal
neutrino mass ordering. In particular it reproduces the best fit values of angles
of all three global fits \cite{King:2013eh} to within one standard deviation.

There have been many attempts to describe the lepton mixing angles
based on the type I see-saw model \cite{Minkowski:1977sc}
combined with sequential dominance (SD) \cite{King:1998jw}
in which the right-handed neutrinos contribute with sequential strength. 
Constrained sequential dominance (CSD) \cite{King:2005bj}
involves the right-handed neutrino mainly responsible for the atmospheric neutrino mass 
having couplings to $(\nu_e, \nu_{\mu}, \nu_{\tau})$ proportional to $(0,1,1)$ and the 
right-handed neutrino mainly responsible for the solar neutrino mass having couplings to $(\nu_e, \nu_{\mu}, \nu_{\tau})$ proportional to $(1,1,-1)$ and it led to TB mixing.
CSD2 \cite{Antusch:2011ic} was proposed to give a non-zero reactor angle 
and is based on the same atmospheric alignment but with 
right-handed neutrino mainly responsible for the solar neutrino mass having couplings to $(\nu_e, \nu_{\mu}, \nu_{\tau})$ proportional to $(1,0,2)$ or $(1,2,0)$ yielding a reactor angle $\theta_{13}\approx 6^\circ$
which unfortunately is too small, although the situation can be rescued by invoking charged lepton corrections
\cite{Antusch:2013wn}.
The CSD3 model in \cite{King:2013iva} involves the
right-handed neutrino mainly responsible for the solar neutrino mass having couplings to $(\nu_e, \nu_{\mu}, \nu_{\tau})$ proportional to $(1,3,1)$ or $(1,1,3)$ with a relative phase $\mp \pi/3$
yielding a reactor angle $\theta_{13}\approx 8.5^\circ$ close to the observed value.
However CSD3 predicts 
approximate TBC mixing with an almost maximal atmospheric mixing angle disfavoured by the latest global fits,
and so it may soon be challenged.

In this paper we shall propose a model based on a new possibility called CSD4
which predicts the above best fit angles in
Eq.\ref{bestfitangles}
of the PMNS lepton mixing matrix and also makes predictions for the physical CP violating phases.
Similar to all SD models, 
the CSD4 model involves effectively two right-handed neutrinos and a normal neutrino mass hierarchy, leading
to $m_1=0$. As in CSD2 and CSD3, the CSD4 model only requires one input parameter,
namely the ratio of neutrino masses which is selected to be $m_2/m_3 \approx 3\lambda /4$,
which is a natural value that one would expect in such models.
Also as in CSD2 and CSD3, once this value is chosen, the entire PMNS mixing matrix is then fixed by the theory 
(up to a discrete choice of phases) with no remaining free parameters. 
In the CSD4 model, 
the right-handed neutrino mainly responsible for the atmospheric neutrino mass 
has couplings to $(\nu_e, \nu_{\mu}, \nu_{\tau})$ proportional to $(0,1,1)$ and the 
right-handed neutrino mainly responsible for the solar neutrino mass has couplings to $(\nu_e, \nu_{\mu}, \nu_{\tau})$ proportional to $(1,4,2)$, with a
relative phase $\eta = -2\pi/5$.
These couplings and phase relation 
were first discovered in \cite{King:2013iva} and shown to lead to lepton mixing angles
in good agreement with the latest global fits, but no model has been proposed based on CSD4. 
The goal of this paper is to show how CSD4 can arise from an $A_4$ family symmetry,
together with additional discrete $Z_5$ and $Z_3$ symmetries,
and to present the first model of leptons along these lines.
This is necessary since it is far from clear whether alignments such as $(1,4,2)$
are possible to achieve within a realistic model.
The CSD4 model presented here predicts the best fit PMNS angles 
in Eq.\ref{bestfitangles} 
with the distinctive prediction for the oscillation phase $\delta \approx 106^\circ$.

\section{A minimal predictive $A_4 $ model of leptons }
In this section we outline a supersymmetric (SUSY) $A_4$ model of leptons with CSD4
along the lines of the $A_4$ models of leptons discussed in 
\cite{Antusch:2011ic,King:2011ab}. The basic idea is that the three families of lepton doublets
$L$ form a triplet of $A_4$ while the right-handed charged leptons
$e^c, \mu^c, \tau^c$, right-handed neutrinos $N_{\rm atm},N_{\rm sol}$ 
and the two Higgs doublets $H_1,H_2$ required by SUSY are all singlets of $A_4$.
In addition the model employs an additional $Z_3^{\theta}$ family symmetry
in order to account for the charged lepton mass hierarchy.
 
The vacuum alignment that is required for the model is discussed in Appendix A.
In Table~\ref{tab-A4} we have displayed the symmetries and superfields relevant for the Yukawa sector only.
In Appendix A the transformation properties 
of the remaining superfields under $Z_3^{l}\times Z_5^{\nu_i}$ 
responsible for vacuum alignment
is discussed and are consistent with the charges shown in Table~\ref{tab-A4},
where we have written $\phi_{\rm atm}\equiv \varphi_{\nu_3}$ and $\phi_{\rm sol}\equiv \varphi_{\nu_4}$
and hence $Z_5^{\rm atm}\equiv Z_5^{\nu_3}$ and $Z_5^{\rm sol}\equiv Z_5^{\nu_4}$.

%Experience with the models in \cite{Antusch:2011ic,King:2011ab} shows
%that a fully realistic model along these lines will be rather involved, and it is our intension
%here only to indicate how CSD3 could in principle be implemented within a realistic model.
%Since this paper is already quite long, we shall therefore defer a detailed discussion
%of such a model, including all the extra auxiliary symmetries and messenger sector that 
%are typically required \cite{Antusch:2011ic,King:2011ab}, to a future study.
%We are not pretending that constructing a realistic model along these lines is easy.

The charged lepton sector of the model employs the $A_4$ triplet flavons $\varphi_e,  \varphi_{\mu}, \varphi_{\tau}$
whose alignment is discussed in Appendix A.
With the lepton symmetries in the upper left of Table~\ref{tab-A4} 
we may enforce the following charged lepton Yukawa superpotential at leading order
\begin{equation}
\mathcal{W}^e_{\text{Yuk}} \sim \frac{1}{\Lambda}H_1 (\varphi_{\tau} \cdot L )\tau^c 
+   \frac{1}{\Lambda^2}  \theta H_1 (\varphi_{\mu}\cdot L) \mu^c +
\frac{1}{\Lambda^3}  \theta^2 H_1( \varphi_e \cdot L) e^c,
\end{equation}
which give the charged lepton Yukawa couplings after the flavons develop their
vevs. $\Lambda$ is a generic messenger mass scale, but in a renormalisable model
the messengers scales may differ. The charged lepton symmetries include 
three lepton flavour symmetries $Z_3^{e, \mu , \tau}$ under which 
$\varphi_e,  \varphi_{\mu}, \varphi_{\tau}$ and $e^c, \mu^c , \tau^c$ 
transform respectively 
as $\omega$ and $\omega^2$, together with a lepton family symmetry $Z_3^{\theta}$ under which 
$e^c, \mu^c , \tau^c$ transform as $\omega , \omega^2 , 1$ respectively
(where $\omega = e^{i2\pi /3}$) with the 
family symmetry breaking flavon $\theta$ transforming as $\omega$ and otherwise being a singlet
under all other symmetries. $H_1$ and $L$ and all other fields 
are singlets under $Z_3^{e, \mu , \tau}$ and $Z_3^{\theta}$.
With these charge assignments the higher order corrections are very suppressed.

The charged lepton Yukawa
matrix is diagonal at leading order due to the alignment of the charged lepton-type flavons in Eq.\ref{a4-align-char0}
(where the driving fields responsible for the alignment in Eq.\ref{a4-align-charged} absorb
the charges under the newly introduced symmetries $Z_3^{e, \mu , \tau}$ and $Z_3^{\theta}$)
and has the form, 
\begin{equation}
\label{Ye}
Y^e = \text{diag}(y_e, y_\mu, y_\tau) \sim \text{diag}(\epsilon^2 , \epsilon , 1)y_\tau
\end{equation}
where we choose $\epsilon \sim \langle \theta \rangle /\Lambda \sim \lambda^2$ in order to generate
the correct order of magnitude charged lepton mass hierarchy, with precise
charged lepton masses also dependent on order one coefficients which we have suppressed here.

\begin{table}
	\centering
$$
\begin{array}{||c||cccccccc||c||ccccccc||}
\hline \hline
&\theta &e^c&\mu^c & \tau^c
&\varphi_e&\varphi_\mu&\varphi_\tau
&H_{1}&L&H_{2}&\phi_{\rm atm}&\phi_{\rm sol}
&N_{\rm atm}&N_{\rm sol}
&\xi_{\rm atm}&\xi_{\rm sol}\\  \hline
\hline
Z_3^{\theta} & \omega &  \omega & \omega^2 &1 & 1  &1
&1&1
&1&1 &1
&1& 1  &1&1&1\\[2mm] 
Z_3^{e} & 1 &  \omega^2   & 1  & 1 &  \omega & 1& 1 & 1 & 1 & 1 & 1 
& 1 & 1 & 1  &1 & 1 \\[2mm] 
Z_3^{\mu} & 1 &  1   &   \omega^2  & 1 & 1 &  \omega & 1 & 1
& 1 & 1  & 1  &  1 & 1  & 1 & 1 & 1 \\[2mm] 
Z_3^{\tau} & 1  & 1  & 1 &   \omega^2 & 1 & 1
&  \omega & 1 & 1&  1 & 1 
& 1 & 1  &  1 & 1& 1\\[2mm] \hline \hline
A_4 & {\bf 1} & {\bf 1} & {\bf 1} & {\bf 1} & {\bf 3} & {\bf 3} 
& {\bf 3}  & {\bf 1}
&{\bf 3} & {\bf 1} & {\bf 3} 
& {\bf 3} & {\bf 1}& {\bf 1}& {\bf 1}& {\bf 1} 
\\[2mm] \hline  \hline
Z_5^{\rm atm} & 1 & 1  & 1 & 1&1&1
&1&1
&1& 1 &\rho^3
&1& \rho^2  &1 &\rho &1\\[2mm] 
Z_5^{\rm sol} & 1 & 1  & 1 & 1&1&1
&1&1
&1& 1 &1
&\rho^3&1 & \rho^2  &1 &\rho \\[2mm] \hline \hline
%
%
%U(1)_R&0&1&1&1&0&0&0&0&1&0&0&0&1&1&0&0 \\ 
%\hline
\end{array}
$$
\caption{\label{tab-A4}Lepton, Higgs and flavon superfields 
and how they transform under the symmetries relevant for the Yukawa sector of the model.
The only non-trivial charged lepton charges are in the upper left of the Table
and the only non-trivial neutrino charges in the lower right of the Table.
Note that the only
the lepton doublets $L$ and $A_4$ symmetry, are common to both 
charged lepton and neutrino sectors and are given near the central column and row.
The Standard Model gauge symmetries and $U(1)_R$ symmetry,
under which all the leptons have a charge of unity while the Higgs and flavons have zero charge,
are not shown in the Table. }
\end{table}

With the neutrino symmetries in the lower right part of Table~\ref{tab-A4} 
we may enforce the following leading order neutrino Yukawa superpotential 
\begin{equation}
\mathcal{W}^{\nu}_{\text{Yuk}} \sim \frac{1}{\Lambda}H_2 (\phi_{\rm atm} \cdot L) N_{\rm atm}  
+ \frac{1}{\Lambda}H_2 (\phi_{\rm sol} \cdot L) N_{\rm sol}  .
\end{equation}
Again the higher order corrections are completely negligible.
The neutrino sector of the model exploits 
the $A_4$ triplet flavons $\phi_{\rm atm}\equiv \varphi_{\nu_3}$, and $\phi_{\rm sol}\equiv \varphi_{\nu_4}$
whose alignment is discussed in Appendix A.

As is typical in models of this kind \cite{Antusch:2011ic,King:2011ab}, the RH neutrinos have no mass terms at the
renormalisable level, but they become massive after 
some $A_4$ singlet flavons $\xi_{\rm atm}$ and 
$\xi_{\rm sol}$ develop their
vevs due to the renormalisable superpotential, 
\begin{equation}
 \mathcal{W}_R \sim \xi_{\rm atm} N_{\rm atm}^2 
 + \xi_{\rm sol} N_{\rm sol}^2  \;.
\end{equation} 
When the 
right-handed neutrino flavons develop their vevs 
$\langle \xi_{\rm atm} \rangle \sim M_A$
together with $\langle \xi_{\rm sol} \rangle \sim M_B$, then the RH neutrino
mass matrix is diagonal as required,
\begin{equation} \label{eq:MR}
 M_R  =  \begin{pmatrix} M_{A} & 0 \\ 0 & M_{B} \end{pmatrix} \;.
\end{equation}
To ensure that the mixed terms are absent at renormalisable order 
we have imposed a right-handed neutrino flavour symmetry
$Z_5^{\rm atm}$ under which $N_{\rm atm}$ and $\xi_{\rm atm}$ transform as $\rho^2$ and $\rho$
(where $\rho = e^{i2\pi /5}$) while $\phi_{\rm atm}$ transforms as  $\rho^3$ 
with all other fields being singlets. We have also imposed a similar symmetry $Z_3^{\rm sol}$
under which the ``solar'' fields transform in an analogous way. 
We remark that these charge assignments are 
consistent with the flavon superpotential in Eq.\ref{a4-flavon-nu}, where we identify 
$\phi_{\rm atm}\equiv \varphi_{\nu_3}$, and $\phi_{\rm sol}\equiv \varphi_{\nu_4}$,
with suitable charges assigned to the driving fields.

With all masses, couplings and messenger scales set approximately equal,
the driving flavon superpotentials 
would predict $|\langle \xi_{\rm atm} \rangle| \approx |\langle \xi_{\rm sol} \rangle |$
and hence approximately equal right-handed neutrino masses $M_A\approx M_B$.
Similarly, from Appendix A with $\phi_{\rm atm}\equiv \varphi_{\nu_3}$ and $\phi_{\rm sol}\equiv \varphi_{\nu_4}$,
we see that 
$\langle  \phi_{\rm atm}^2 \rangle \approx \langle  \phi_{\rm sol}^2 \rangle $
would lead to 
\begin{equation}
\label{Phi8} 
\vev{\phi_{\rm atm}}
=
\frac{1}{\sqrt{2}}  \begin{pmatrix}0 \\ 1 \\ 1\end{pmatrix}v_{\rm atm},
 \qquad
\vev{\phi_{\rm sol}}
= 
\frac{1}{\sqrt{21}} \begin{pmatrix}1 \\ 4 \\ 2\end{pmatrix}v_{\rm sol},
\end{equation}
where $v_{\rm atm}\approx v_{\rm sol}$. 

The above charge assignments allow higher order non-renormalisable mixed terms 
such as 
\begin{equation}
 \Delta \mathcal{W}_R \sim \frac{1}{\Lambda} (\phi_{\rm atm} . \phi_{\rm sol}) N_{\rm atm} N_{\rm sol}  \;,
\end{equation} 
which contribute off-diagonal terms to the right-handed neutrino mass matrix
of a magnitude which depends on the absolute scale of the 
flavon vevs $\vev{\phi_{\rm atm}}$ and $\vev{\phi_{\rm sol}}$ compared to 
$\langle \xi_{\rm atm} \rangle $ and $\langle \xi_{\rm sol} \rangle $.
If all flavon vevs and messenger scales are set equal then these terms are suppressed by $\epsilon \sim \lambda^2$
according to the estimate below Eq.\ref{Ye}, however they may be even more suppressed.
We shall ignore the contribution of such off-diagonal mass terms in the following.

Implementing the see-saw mechanism, the effective neutrino mass matrix has the form,
\beq
\label{seesaw2}
m^{\nu}\sim \frac{v_2^2}{\Lambda^2}\left(\frac{\vev{\phi_{\rm atm}}\vev{\phi_{\rm atm}}^T}{\langle \xi_{\rm atm} \rangle}
+  \frac{\vev{\phi_{\rm sol}}\vev{\phi_{\rm sol}}^T}{\langle \xi_{\rm sol} \rangle}\right),
\eeq
where $v_2=\langle H_2 \rangle$.
Hence it can be parameterised, up to an overall irrelevant phase, as, 
\beq
m^{\nu}
= m_a \begin{pmatrix} 0 & 0 & 0 \\ 0 & 1 & 1 \\ 0 & 1 & 1 \end{pmatrix} 
+ m_a e^{2i\eta}\epsilon_{\nu}\begin{pmatrix} 1 & 4 & 2 \\ 4 & 16 & 8 \\ 2 & 8 & 1  \end{pmatrix}
\label{seesaw3}
\eeq
where $m_a$ and $\epsilon_{\nu}$ are real mass parameters which determine the physical neutrino masses 
$m_3$ and $m_2$ and we written the relative phase difference between the two terms as $2\eta $. 
Using Eq.\ref{Phi8} the see-saw mechanism naturally leads to the 
neutrino mass matrix in Eq.\ref{seesaw3} with $\epsilon_{\nu}\approx 2/21 \approx 0.1$.
Hence the desired value of  $\epsilon_{\nu}\approx 0.06$
assumed below is not unreasonable, and may be achieved for example
by taking $M_B\approx 2M_A$. As discussed in \cite{King:2013iva}, we shall also require a special phase relation
$\eta =-2\pi /5$ in order to achieve our goal of predicting the best fit values of the lepton mixing angles.

The phase difference $\eta =-2\pi /5$ between flavon vevs
can be obtained in the context of spontaneous CP violation from discrete symmetries as discussed in 
\cite{Antusch:2013wn}, and we shall follow the strategy outlined there.
The basic idea is to impose CP conservation on the theory so that all couplings and masses are real.
Note that the $A_4$ assignments in Table~\ref{tab-A4} do not involve the complex
singlets $1',1''$ or any complex Clebsch-Gordan coefficients so that the definition of CP is 
straightforward in this model and hence CP may be defined in different ways which are equivalent for our purposes 
(see \cite{Antusch:2013wn} for a discussion of this point).
The CP symmetry is broken in a discrete way by the form of the superpotential terms.
We shall follow \cite{Antusch:2013wn} and suppose that the flavon vevs  
$\vev{\phi_{\rm atm}}$ and $\vev{\phi_{\rm sol}}$ to be real with the phase $\eta$ in Eq.\ref{seesaw3}
originating from the solar right-handed neutrino mass due to the flavon vev 
$\langle \xi_{\rm sol} \rangle \sim M_Be^{4i\pi/5}$
having a complex phase of $4\pi /5$, while the flavon vev $\langle \xi_{\rm atm} \rangle \sim M_A$ is real and positive. This can be arranged if the right-handed neutrino flavon vevs arise from the superpotential,
\be
W_{A_4}^{\mathrm{flavon},R} = g P\left(\frac{\xi_{\rm atm}^5}{\Lambda^3}  -M^2\right) + 
g'P' \left(\frac{\xi_{\rm sol}^5}{\Lambda'^3}  - M'^2\right) ,
\label{Rflavon}
\ee
where, as in \cite{Antusch:2013wn}, the driving singlet fields $P,P'$ denote linear combinations 
of identical singlets and all couplings and masses are real due to CP conservation.
The F-term conditions from Eq.\ref{Rflavon} are,
\begin{equation}
 \left| \frac{\langle \xi_{\rm atm} \rangle^5}{\Lambda^3} - M^2\right|^2 
 = \left| \frac{\langle \xi_{\rm sol} \rangle^5}{\Lambda'^3} - M'^2\right|^2 = 0 .
\end{equation} 
These are satisfied by $\langle \xi_{\rm atm} \rangle = |(\Lambda^3 M^2)^{1/5}|$ and 
$\langle \xi_{\rm sol} \rangle =  |(\Lambda'^3M'^2)^{1/5}|e^{4i\pi/5}$
where we arbitrarily select the phases to be 
zero and $4\pi /5$ from amongst a discrete set of possible choices
in each case. More generally we require a phase difference of $4\pi /5$ since the overall phase is not physically
relevant, which would happen one in five times by chance.
In the basis where the right-handed neutrino masses are real and positive
this is equivalent to having a phase difference $\eta  = - 2\pi /5$ between flavon vevs in Eq.\ref{Phi8}
according to the see-saw result in Eq.\ref{seesaw2}.

Similarly the flavons appearing in Eqs.\ref{Delta-a4-align-charged} each have
a discrete choice of phases. The charged lepton flavons $\varphi_l$ may take any phases
since such phases are unphysical. In fact the only physically significant flavon phases from the previous subsection
are those of 
$\phi_{\rm atm}\equiv \varphi_{\nu_3}$, and $\phi_{\rm sol}\equiv \varphi_{\nu_4}$
whose phases are selected to be equal.
As before, this would occur one in five times by chance.

We emphasise that, with the alignments including the phase $\eta$
fixed, the neutrino mass matrix is completely determined by only two
parameters, namely an overall mass scale $m_a$, which may be taken
to fix the atmospheric neutrino mass $m_3=0.048-0.051$ eV, the ratio 
of input masses $\epsilon_{\nu}$, which may be taken to fix the solar to atmospheric neutrino mass
ratio $m_2/m_3 = 0.17-0.18$. In particular the entire PMNS mixing matrix and all the 
parameters therein are then predicted as a function
of $m_2/m_3$ controlled by the only remaining
parameter $\epsilon_{\nu}$. 
In Table~\ref{predictions142} we show the predictions for 
CSD4 as a function of $\epsilon_{\nu}$ and hence $m_2/m_3$.

We remark that an accuracy of one degree in the angles is all that can be expected due to purely theoretical
corrections in a realistic model due to renormalisation group running \cite{Boudjemaa:2008jf} and
canonical normalisation corrections \cite{Antusch:2007ib}.
In addition, there may be small contributions from a heavy third right-handed neutrino \cite{Antusch:2010tf}
which can affect the results.

\begin{table}
	\centering
		\begin{tabular}{|c||c|c|c|c|c|c|}
			\hline
			 $\epsilon_{\nu}$ & $m_2/m_3$ &  $\theta_{12}$ 
			 & $\theta_{13}$  & $\theta_{23}$  & $\delta$  & $\beta$ \\ \hline \hline
			0.057 &0.166 &34.2$^{\circ}$ & 9.0$^{\circ}$ & 40.8$^{\circ}$ & 107$^{\circ}$ & -84$^{\circ}$\\ \hline   
			0.058 &0.170 &34.2$^{\circ}$ & 9.2$^{\circ}$ & 40.9$^{\circ}$ & 107$^{\circ}$ & -83$^{\circ}$\\ \hline 
			0.059 &0.174 &34.1$^{\circ}$ & 9.4$^{\circ}$ & 41.0$^{\circ}$ & 106$^{\circ}$ & -82$^{\circ}$\\ \hline
			0.060 &0.177 &34.1$^{\circ}$ & 9.6$^{\circ}$ & 41.1$^{\circ}$ & 105$^{\circ}$ & -80$^{\circ}$\\ \hline  
			0.061 &0.181 &34.1$^{\circ}$ & 9.7$^{\circ}$ & 41.3$^{\circ}$ & 104$^{\circ}$ & -79$^{\circ}$\\ \hline     
				\end{tabular} 
			\caption{The predictions for PMNS parameters and $m_2/m_3$
			arising from CSD4 as a function of 
			$\epsilon_{\nu}$. Note that these predictions assume $\eta = -2\pi/5$.
			The predictions are obtained numerically using the Mixing Parameter Tools
package based on \cite{Antusch:2005gp}.
The leading order analytic results are not reliable as discussed in Appendix B.
			}
		\label{predictions142}
\end{table}

As in the case of CSD2, the neutrino mass matrix implies the TM$_1$ 
mixing form \cite{Lam:2006wm} where the first column of the PMNS matrix is proportional to
$(2,-1,1)^T$. The reason is simply that $\langle \varphi_{\nu_1} \rangle \propto (2,-1,1)^T$
is an eigenvector of $m^{\nu}$ in Eq.\ref{seesaw3} with a zero
eigenvalue corresponding to the first neutrino mass $m_1$ being zero. The reason for
this is that $m^{\nu}$ in Eq.\ref{seesaw3} is a sum of two terms, the
first being proportional to $AA^T\propto \langle \varphi_{\nu_3} \rangle \langle \varphi_{\nu_3} \rangle^T$  and
the second being proportional to $BB^T\propto \langle \varphi_{\nu_4} \rangle \langle \varphi_{\nu_4} \rangle^T$.  
Since $\langle \varphi_{\nu_1} \rangle \propto (2,-1,1)^T$ is orthogonal to both
$\langle \varphi_{\nu_3} \rangle$ and $\langle \varphi_{\nu_4} \rangle$ it is then clearly annihilated by
the neutrino mass matrix, i.e. it is an eigenvector with zero
eigenvalue. Therefore we immediately expect $m^{\nu}$ in Eq.\ref{seesaw3}
to be diagonalised by the TM$_1$ mixing matrix \cite{Lam:2006wm} where
the first column is proportional to $\langle \varphi_{\nu_1} \rangle \propto (2,-1,1)^T$. 
Therefore we already know that CSD4
must lead to TM$_1$ mixing exactly to all orders according to this general argument.

Exact  TM$_1$ mixing angle and phase relations are obtained by
equating moduli of PMNS elements to those of the first column of the TB mixing matrix:
\bea
c_{12}c_{13}=\sqrt{\frac{{2}}{{3}}} ,\label{a1} \\
|c_{23}s_{12}+s_{13}s_{23}c_{12}e^{i\delta}|=\frac{1}{\sqrt{6}} , \label{a2}\\
|s_{23}s_{12}-s_{13}c_{23}c_{12}e^{i\delta}|=\frac{1}{\sqrt{6}} . \label{a3}
\eea
From Eq.\ref{a1} we see that TM$_1$ mixing approximately preserves the successful TB mixing for the solar
mixing angle $\theta_{12}\approx 35^\circ$ as the correction due to a non-zero
but relatively small reactor angle is of second order. 
While general TM$_1$ mixing involves an undetermined reactor
angle $\theta_{13}$, we emphasise that CSD4 fixes this reactor angle.
The approximate leading order result is
\be
\theta_{13}  \approx  \frac{4}{3\sqrt{2}}\frac{m_2}{m_3}.
\label{1333}
\ee
However the leading order results are not highly accurate and numerically the prediction for the reactor angle
depends on the phase $\eta$. For $\eta = -2\pi/5$ the reactor angle is in the correct range as shown
in Table~\ref{predictions142}.

In an approximate linear form, the relations in Eq.\ref{a1}-\ref{a3} imply the atmospheric sum rule relation,
\beq
\theta_{23}\approx 45^\circ +\sqrt{2}\theta_{13}\cos \delta .
\label{atmsum}
\eeq
For $\eta = -2\pi/5$ the predictions shown in Table~\ref{predictions142} for the small deviations
of the atmospheric angle from maximality are well described by the sum rule in Eq.\ref{atmsum}.
In the present model this sum rule is satisfied by particular predicted values of angles and CP phase
which only depend on the neutrino mass ratio $m_2/m_3$. Over the successful range of $m_2/m_3$ we predict
CP violation with $\delta \approx  106^\circ$ and $\theta_{23}\approx 41^\circ$ which satisfy the sum rule.
Note that according to this sum rule, non-maximal atmospheric mixing is linked to non-maximal CP violation.

\section{Conclusions}
\label{conclusion}
There is long history of attempts to explain the neutrino mixing angles starting from the 
type I see-saw mechanism and using SD, first using CSD to account for TB mixing, then
using CSD2 to obtain a small reactor angle before going to CSD3 where the correct reactor
angle can be reproduced along with maximal atmospheric mixing. 
We have discussed a minimal predictive see-saw model based on CSD4 
in which the right-handed neutrino mainly responsible for the atmospheric neutrino mass has couplings to $(\nu_e, \nu_{\mu}, \nu_{\tau})$ proportional to $(0,1,1)$ and the 
right-handed neutrino mainly responsible for the solar neutrino mass has couplings to $(\nu_e, \nu_{\mu}, \nu_{\tau})$ proportional to $(1,4,2)$, with a
relative phase $\eta = -2\pi/5$.
We have shown how these patterns of couplings and phase could arise from an $A_4$ family symmetry
model of leptons. 

We remark that the type of model presented here is referred to as ``indirect'' according to the classification
scheme of models in \cite{King:2013eh}, meaning that the family symmetry is completely broken
by flavons and its only purpose is to generate the desired vacuum alignments. By contrast, the ``direct'' models 
where the symmetries of the neutrino and charged lepton mass matrices is identified as
a subgroup of the family symmetry, requires
rather large family symmetry groups in order to account for the reactor angle \cite{King:2013vna}. 
It is possible to have ``semi-direct'' models, 
either at leading order or emerging due to higher order corrections 
\cite{King:2013eh}, but these are inherently less predictive. In the light of the observed reactor angle, 
``indirect models'' therefore 
offer the prospect of full predictivity at the leading order from a small family symmetry group.
Spontaneous CP violation seems to be an important ingredient in the ``indirect'' approach since a particular phase
relation between flavons a crucial requirement.

The particular indirect model presented here, in which CSD4 emerges from an $A_4$ family symmetry,
offers a highly predictive framework involving only one free parameter 
which is used to fix the neutrino mass ratio
$m_2/m_3$, together with an overall neutrino mass scale which is used to fix the atmospheric neutrino mass $m_3$. 
Remarkably, the model then predicts the PMNS angles 
$\theta_{12}\approx 34^{\circ}, \theta_{23}\approx 41^{\circ}, \theta_{13}\approx 9.5^{\circ}$,
which exactly coincide with the current best fit values for a normal neutrino mass hierarchy,
together with the distinctive prediction for the CP violating oscillation phase $\delta \approx 106^\circ$.
These predictions will surely be tested by current and planned high precision neutrino oscillation experiments.

\section*{Acknowledgements}
SFK would like to thank A. Merle for help with MPT,
Christoph Luhn and Stefan Antusch for discussions
and A. Kusenko and T. Yanagida and the IPMU for hospitality and support.
SFK also acknowledges partial support 
from the STFC Consolidated ST/J000396/1 and EU ITN grants UNILHC 237920 and INVISIBLES 289442 .

\appendix
\section{Vacuum alignment}

In this appendix we shall discuss how to achieve the following vacuum alignment,
\begin{equation}
\label{Phi7} 
\frac{\vev{\phi_{\rm atm}}}{\Lambda}
\propto  \begin{pmatrix}0 \\ 1 \\ 1\end{pmatrix},
 \qquad
\frac{\vev{\phi_{\rm sol}}}{\Lambda}
\propto  
 \begin{pmatrix}1 \\ 4 \\ 2\end{pmatrix},
\end{equation}
which we refer to as CSD4.

The vacuum alignments associated with TB mixing have been very well studied.
Here we shall focus on the family symmetry $A_4$ as it is the smallest non-Abelian
finite group with an irreducible triplet representation. 
The generators of the $A_4$ group,
can be written as $S$ and $T$ with $S^2=T^3=(ST)^3=\mathcal{I}$.
$A_4$ has four irreducible representations, three singlets
$1,~1^\prime$ and $1^{\prime \prime}$ and one triplet. 
The products of singlets are:
\begin{equation}\begin{array}{llll}
1\otimes1=1&1^\prime\otimes1^{\prime\prime}=1&1^{\prime}\otimes1^{\prime}=1^{\prime\prime}%\\
&1^{\prime\prime}\otimes1^{\prime\prime}=1^\prime.
\end{array}
\end{equation}
We work in the basis~\cite{A4-refs}, 
\begin{equation}\label{eq:ST}
S=\left(
\begin{array}{ccc}
1&0&0\\
0&-1&0\\
0&0&-1\\
\end{array}
\right), \ \ \ \ 
T=\left(
\begin{array}{ccc}
0&1&0\\
0&0&1\\
1&0&0\\
\end{array}
\right)\,.
\end{equation}
In this basis one has the following Clebsch rules for the multiplication of two triplets, 
\begin{equation}\label{pr}
\begin{array}{lll}
(ab)_1&=&a_1b_1+a_2b_2+a_3b_3\,;\\
(ab)_{1'}&=&a_1b_1+\omega a_2b_2+\omega^2a_3b_3\,;\\
(ab)_{1''}&=&a_1b_1+\omega^2 a_2b_2+\omega a_3b_3\,;\\
(ab)_{3_1}&=&(a_2b_3,a_3b_1,a_1b_2)\,;\\
(ab)_{3_2}&=&(a_3b_2,a_1b_3,a_2b_1)\,,
\end{array}
\end{equation}
where $\omega^3=1$, $a=(a_1,a_2,a_3)$ and $b=(b_1,b_2,b_3)$.

Following the methods of \cite{King:2011ab} it is straightforward to obtain the vacuum 
alignments for charged lepton flavon alignments suitable for a diagonal charged lepton mass matrix.
The charged lepton flavon alignments used to generate a diagonal charged lepton mass
matrix are obtained from the renormalisable superpotential \cite{King:2011ab},
\be
W_{A_4}^{\mathrm{flavon},\ell}~\sim~
A_e \varphi_e \varphi_e + A_\mu \varphi_\mu \varphi_\mu 
+ A_\tau \varphi_\tau \varphi_\tau
+O_{e\mu} \varphi_e  \varphi_\mu+ O_{e\tau} \varphi_e  \varphi_\tau
+ O_{\mu\tau} \varphi_\mu  \varphi_\tau \ .  \label{a4-align-charged}
\ee
The triplet driving fields $A_{e,\mu,\tau}$ give rise to flavon alignments
$\langle \varphi_{e,\mu,\tau} \rangle$
with two zero components, and the singlet driving fields $O_{ij}$ require
orthogonality among the three flavon VEVs so that we arrive at the vacuum
structure \cite{King:2011ab},
\be
\langle \varphi_e \rangle =v_e 
\begin{pmatrix} 1\\0\\0 \end{pmatrix}  \ , \qquad
\langle \varphi_\mu \rangle =v_\mu 
\begin{pmatrix} 0\\1\\0 \end{pmatrix} \ , \qquad
\langle \varphi_\tau \rangle = v_\tau 
\begin{pmatrix} 0\\0\\1 \end{pmatrix} 
 \ .\label{a4-align-char0}
\ee
%Since only wish to modify the neutrino sector flavons we shall take the 
%charged lepton alignments as given, referring the reader to \cite{King:2011ab}
%for details of how the charged lepton flavon alignments are obtained.
Of more interest to us in this paper are the new neutrino flavon alignments.
The starting point for the discussion is the usual standard 
TB neutrino flavon alignments proportional to the respective columns of the TB mixing matrix,
\be
\langle \varphi_{\nu_1} \rangle = 
v_{\nu_1} \begin{pmatrix} 
  2\\-1\\1 \end{pmatrix}  , 
\qquad
\langle \varphi_{\nu_2} \rangle = 
v_{\nu_2} \begin{pmatrix} 
  1\\1\\-1 \end{pmatrix}  , 
\qquad
\langle \varphi_{\nu_3} \rangle =  
v_{\nu_3} \begin{pmatrix} 0\\1\\1 \end{pmatrix} .
\label{a4-align-nu}
\ee
We will also employ the alternative TB alignments which are related by phase redefinitions,
\be
\langle \varphi_{\nu'_1} \rangle = 
v'_{\nu_1} \begin{pmatrix} 
  2\\1\\-1 \end{pmatrix}  , 
\qquad
\langle \varphi_{\nu'_2} \rangle = 
v_{\nu'_2} \begin{pmatrix} 
  1\\-1\\1 \end{pmatrix}  , 
\qquad
\langle \varphi_{\nu'_3} \rangle =  
v_{\nu'_3} \begin{pmatrix} 0\\-1\\-1 \end{pmatrix} .
\label{a4-align-nu-primed}
\ee
In the remainder of this subsection we shall show how to obtain the neutrino flavon alignments including the new
alignment,
\be
\langle \varphi_{\nu_4} \rangle = 
v_{\nu_4} \begin{pmatrix} 
  1\\4\\2 \end{pmatrix} ,
\label{a4-align-nu-4}
\ee
which corresponds to the CSD4 solar flavon alignment in Eq.\ref{Phi7}.
We shall identify 
$\phi_{\rm atm}\equiv \varphi_{\nu_3}$, and $\phi_{\rm sol}\equiv \varphi_{\nu_4}$.
The renormalisable superpotential involving the driving fields
necessary for aligning the neutrino-type flavons is given as
\bea
W_{A_4}^{\mathrm{flavon},\nu}&=& 
A_{\nu_2} (g_1 \varphi_{\nu_2}\varphi_{\nu_2}  
+ g_2 \varphi_{\nu_2} \xi_{\nu_2}  )
+A_{\nu'_2} (g'_1 \varphi_{\nu'_2}\varphi_{\nu'_2}  
+ g'_2 \varphi_{\nu'_2} \xi_{\nu'_2}  )
\label{a4-flavon-nu}
\\[2mm] &&
+ \,O_{e\nu_3} g_3 \varphi_e \varphi_{\nu_3}  +  O_{\nu_2 \nu_3} g_4 \varphi_{\nu_2} \varphi_{\nu_3} 
+  O_{\nu_1 \nu_2} g_5 \varphi_{\nu_1} \varphi_{\nu_2} 
+  O_{\nu_1 \nu_3} g_6 \varphi_{\nu_1} \varphi_{\nu_3} 
\notag \\[2mm] &&
+ \,O_{e\nu'_3} g'_3 \varphi_e \varphi_{\nu'_3}  +  O_{\nu'_2 \nu'_3} g'_4 \varphi_{\nu'_2} \varphi_{\nu'_3} 
+  O_{\nu'_1 \nu'_2} g'_5 \varphi_{\nu'_1} \varphi_{\nu'_2} 
+  O_{\nu'_1 \nu'_3} g'_6 \varphi_{\nu'_1} \varphi_{\nu'_3} 
\notag \\[2mm] &&
+ \,O_{\mu \nu_5} g_7 \varphi_{\mu} \varphi_{\nu_5} +  O_{\nu'_1 \nu_5} g_8 \varphi_{\nu'_1} \varphi_{\nu_5}
+ O_{\mu {\nu_6}} g_9 \varphi_{\mu} \varphi_{\nu_6} +  O_{{\nu_5} {\nu_6}} g_{10}\varphi_{\nu_5} \varphi_{\nu_6}
\notag \\[2mm] &&
+\,O_{{\nu_6} \nu_4} g_{11} \varphi_{\nu_6} \varphi_{\nu_4}
+\,O_{\nu_1 \nu_4} g_{12} \varphi_{\nu_1} \varphi_{\nu_4},
\notag
\eea
where $A_{\nu_2}$ is a triplet driving field and $O_{ij}$ are singlet driving fields whose
F-terms lead to orthogonality relations between the accompanying flavon fields.
Here $g_i$ are dimensionless coupling constants.
The first line of Eq.~\eqref{a4-flavon-nu} produces the vacuum alignment 
$\langle \varphi_{\nu_2} \rangle \propto (1,1,-1)^T$ of Eq.~\eqref{a4-align-nu}
and
$\langle \varphi_{\nu'_2} \rangle \propto (1,-1,1)^T$ of Eq.~\eqref{a4-align-nu-primed}
as can be seen from the $F$-term conditions
\footnote{We remark that the
  general alignment derived from these $F$-term conditions is $\langle
  \varphi_{\nu_2} \rangle \propto (\pm1 ,\pm 1,\pm1)^T$. One can, however,
  show that all of them are equivalent up to phase redefinitions. 
  Note that $(1,1,-1)$ is related to permutations of the minus sign as well as to 
  $(-1,-1,-1)$ by $A_4$ transformations. The other four choices can be obtained from these by simply multiplying an overall phase (which would also change the sign of the $\xi_{\nu_2}$ vev.) } 
\be
2 g_1 \begin{pmatrix}
\langle \varphi_{\nu_2} \rangle_2  \langle \varphi_{\nu_2} \rangle_3 \\
\langle \varphi_{\nu_2} \rangle_3  \langle \varphi_{\nu_2} \rangle_1 \\
\langle \varphi_{\nu_2} \rangle_1  \langle \varphi_{\nu_2} \rangle_2 
\end{pmatrix}
+ g_2 \langle \xi_{\nu_2}  \rangle 
\begin{pmatrix}
\langle \varphi_{\nu_2} \rangle_1\\
\langle \varphi_{\nu_2} \rangle_2\\
\langle \varphi_{\nu_2} \rangle_3
\end{pmatrix} ~=~ 
\begin{pmatrix} 0\\0\\0
\end{pmatrix} . 
\ee
plus similar conditions involving the primed flavons.
The first two terms in the second line of Eq.~\eqref{a4-flavon-nu} give rise to
orthogonality conditions which uniquely fix the alignment 
$\langle \varphi_{\nu_3} \rangle \propto (0,1,1)^T$ of Eq.~\eqref{a4-align-nu},
\bea
\langle \varphi_e \rangle^T \cdot \langle \varphi_{\nu_3} \rangle \,=\, 
\langle \varphi_{\nu_2} \rangle^T \cdot \langle \varphi_{\nu_3} \rangle \,=\, 0 
\quad 
&\rightarrow&
\quad
\langle \varphi_{\nu_3} \rangle \,\propto\, \begin{pmatrix} 0\\1\\1\end{pmatrix} \ .
\eea

The last two terms in the second line of Eq.~\eqref{a4-flavon-nu} give rise to
orthogonality conditions which uniquely fix the alignment 
$\langle \varphi_{\nu_1} \rangle \propto (2,-1,1)^T$ of Eq.~\eqref{a4-align-nu},
\bea
\langle \varphi_{\nu_1} \rangle^T \cdot \langle \varphi_{\nu_2} \rangle \,=\, 
\langle \varphi_{\nu_1} \rangle^T \cdot \langle \varphi_{\nu_3} \rangle \,=\, 0 
\quad 
&\rightarrow&
\quad
\langle \varphi_{\nu_1} \rangle \,\propto\, \begin{pmatrix} 2\\-1\\1\end{pmatrix} \ .
\eea

Similarly the terms in the third line of Eq.~\eqref{a4-flavon-nu} give rise to
orthogonality conditions which fix the alternative TB alignments in
Eq.\ref{a4-align-nu-primed} corresponding to a different choice of phases.

The terms in the fourth line of Eq.~\eqref{a4-flavon-nu} give rise to
orthogonality conditions which fix the alignments
of the auxiliary
flavon fields $\varphi_{\nu_5}$ and $\varphi_{\nu_6}$,
\bea
\langle \varphi_{\mu} \rangle^T \cdot \langle \varphi_{\nu_5} \rangle \,=\, 
\langle \varphi_{\nu'_1} \rangle^T \cdot \langle \varphi_{\nu_5} \rangle \,=\, 0 
\quad 
&\rightarrow&
\quad
\langle \varphi_{\nu_5} \rangle \,\propto\, \begin{pmatrix} 1\\0\\2\end{pmatrix} \ , \\
\langle \varphi_{\mu} \rangle^T \cdot \langle \varphi_{\nu_6} \rangle \,=\, 
\langle \varphi_{\nu_5} \rangle^T \cdot \langle \varphi_{\nu_6} \rangle \,=\, 0 
\quad 
&\rightarrow&
\quad
\langle \varphi_{\nu_6} \rangle \,\propto\, \begin{pmatrix} 2\\0\\-1\end{pmatrix} \ .
\eea

The neutrino-type flavon of interest labelled as $\varphi_{\nu_4}$ gets aligned
by the remaining terms in the fifth line of Eq.~\eqref{a4-flavon-nu},
leading to the desired alignment in Eq.\ref{a4-align-nu-4},
\bea
\langle \varphi_{\nu_1} \rangle^T \cdot \langle \varphi_{\nu_4} \rangle \,=\, 
\langle \varphi_{\nu_6} \rangle^T \cdot \langle \varphi_{\nu_4} \rangle \,=\, 0 
\quad 
&\rightarrow&
\quad
\langle \varphi_{\nu_4} \rangle \,\propto\, \begin{pmatrix} 1\\4\\2\end{pmatrix} \ .
\eea

So far we have only shown how to align the flavon vevs and have not enforced them to be non-zero.
In order to do this we shall introduce the 
additional non-renormalisable superpotential terms which include,
\bea
\Delta W_{A_4}^{\mathrm{flavon},\ell}&\sim &  \sum_{l=e,\mu , \tau} 
\frac{P}{\Lambda} \left( (\varphi_{l} \cdot \varphi_{l} ) \rho_{l} - M^3 \right)
+ 
P( \frac{\rho_{l}^3 }{\Lambda}  - M^2), \label{Delta-a4-align-charged}
     \\
\Delta W_{A_4}^{\mathrm{flavon},\nu} & \sim &   \sum_{i=1}^6 \frac{P}{\Lambda} \left( (\varphi_{\nu_i}) \cdot \varphi_{\nu_i} ) \rho_{\nu_i} - M^3 \right)  
+P( \frac{\rho_{\nu_i} ^5}{\Lambda^3}  - M^2), \label{Delta-a4-flavon-nu}
\eea
where, as in \cite{Antusch:2011ic}, the driving singlet fields $P$ denote linear combinations 
of identical singlets and we have introduced explicit masses $M$
to drive the non-zero vevs, as well as the messenger scales denoted as $\Lambda$.
We have also introduced $A_4$ singlets $\rho_{l}$ and $\rho_{\nu_i}$ whose vevs are driven by the 
F-terms of the singlets $P$ in the second terms in Eqs.\ref{Delta-a4-align-charged}
and \ref{Delta-a4-flavon-nu}. These singlet vevs enter 
the first terms in Eqs.\ref{Delta-a4-align-charged}
and \ref{Delta-a4-flavon-nu} which drive the vevs of the triplet flavons.

The flavons and driving fields introduced in this Appendix
transform under $Z_3^{l}\times Z_5^{\nu_i}$ symmetries
whose purpose is to allow only the terms
in Eqs.\ref{a4-align-charged}, \ref{a4-flavon-nu}
and \ref{Delta-a4-align-charged} and forbid all other terms.
The superfields $\varphi_{\nu_i}$ transform
under $Z_5^{\nu_i}$ as $\rho^3$
(where $\rho = e^{i2\pi /5}$) and are singlets under all other discrete symmetries.
The superfields $\rho_{\nu_i}$ transform
under $Z_5^{\nu_i}$ as $\rho^4$ and are singlets under all other discrete symmetries. 
Any superfield with a single subscript $l$ transforms under $Z_3^{l}$ as $\omega$ 
(where $\omega = e^{i2\pi /3}$) and is a singlet under all other discrete symmetries.
The orthogonality driving superfields $O_{ij}$ with two subscripts transform under $Z_3^{l}\times Z_5^{\nu_i}$
in such a way as to allow the terms in Eqs.\ref{a4-align-charged}, \ref{a4-flavon-nu}.
For example the  $O_{l {\nu_i}}$ driving fields transform under $Z_3^{l}\times Z_5^{\nu_i}$
as $(\omega^2,\rho^2)$. In addition driving superfields are assigned a charge of two 
while flavon superfields have zero charge under a $U(1)_R$ symmetry.

\section{Leading Order Analytic Results}
For the case of 
atmospheric alignments of the form $(0,z_1,1)$ and
solar alignments of the form $(1,z_2,z_3)$,
the leading order analytic results in \cite{King:1998jw,King:2013iva} give,
\bea
\tan \theta_{23} & \approx &|z_1| \label{231}\\
\cot \theta_{12} & \approx &
c_{23}|z_2|\cos\left( \eta_2-\frac{\beta}{2} \right)-s_{23}|z_3|\cos \left( \eta_3-\frac{\beta}{2}\right) \label{121}\\
\theta_{13} & \approx & \frac{m_2}{m_3}s_{12}^2c_{23}\left||z_3|  +|z_2|\tan \theta_{23}
e^{i(\eta_2 -\eta_3)}     \right|
\label{131}
\eea
where $ \eta_2 = \arg(z_2/z_1)$ and  $\eta_3 = \arg(z_3)$, while $\beta$ is a Majorana phase.
%\bea
%\tan \theta_{23} & \approx & 1 \label{2311}\\
%\cot \theta_{12} & \approx & \frac{1}{\sqrt{2}}\cos\left( \eta -\frac{\beta}{2} \right)
%(|z_2|-|z_3|) \label{1211}\\
%\theta_{13} & \approx & \frac{m_2}{m_3}\frac{1}{3\sqrt{2}}\left||z_3|  +|z_2|   \right|
%\label{1311}
%\eea
%where we have assumed $s_{12}^3\approx 1/3$.
With $z_1=1$ and arbitrarily
assuming $\beta = 0$ and real phases $\pm 1$ 
associated with $\eta_2$ and $\eta_3$ one finds the relations
\bea
\tan \theta_{23} & \approx & 1 \label{2311}\\
\cot \theta_{12} & \approx &
\frac{1}{\sqrt{2}}|z_2 - z_3| \label{1211}\\
\theta_{13} & \approx & \frac{m_2}{m_3}\frac{1}{3\sqrt{2}}\left|z_3  +z_2   \right| .
\label{1311}
\eea
We should say immediately that the assumption $\beta =0$ is not justified so these results
can be at best suggestive.
With this caveat, we 
note that approximately trimaximal solar mixing $\cot \theta_{12}\approx \sqrt{2}$ results from the 
general condition
$|z_2 - z_3| =2$ which is satisfied by all the proposed forms of CSD
\footnote{I would like to thank Stefan Antusch (private communications) for 
emphasising the condition $|z_2 - z_3| =2$.}.
Moreover CSD with $z_2=1$, $z_3=-1$ leads to
$\theta_{13}  \approx  0$,
CSD2 with $z_2=2$, $z_3=0$ leads to $\theta_{13}  \approx  \frac{m_2}{m_3}\frac{\sqrt{2}}{3}$,
CSD3 with $z_2=3$, $z_3=1$ leads to 
$\theta_{13}  \approx  \frac{m_2}{m_3}\frac{4}{3\sqrt{2}}$,
CSD4 with $z_2=4$, $z_3=2$ leads to 
$\theta_{13}  \approx  \frac{m_2}{m_3}\sqrt{2}$.

Although the above leading order results provide a qualitative understanding
of the results obtained for CSD, CSD2, CSD3 and CSD4, 
they have large corrections of order $m_2/m_3$, much larger than the errors in the global fits
and so do not give reliable predictions.
In addition there is a strong dependence
on the phase difference between the solar and atmospheric alignments
which these results ignore. Moreover the phase $\eta$ does not appear in the leading order
formula for $\theta_{13}$, but in practice the reactor angle depends strongly on $\eta$,
as discussed in \cite{King:2013iva}. On the other hand, while the phases do appear in the solar angle formula,
we have arbitrarily and incorrectly assumed $\beta =0$. 

In summary, the leading order results, while providing a qualitative understanding, are 
quantitatively unreliable
and cannot be used to estimate the mixing angles to the required accuracy.
The general analysis as performed in \cite{King:2013iva} did not rely on the leading order results in any way
and was not inspired by them. Starting from an exact master formula, the analysis \cite{King:2013iva}
determined from first principles not only the 
moduli $z_i$ but also the phases $\eta_i$ which are required for a proper
definition of any new type of CSD. For example CSD4 
with solar alignment $(1,4,2)$ is only properly defined 
once the phases $\eta_2 = \eta_3 = -2\pi /5$ are specified.
In retrospect, it might seem rather fortuitous that 
the condition $|z_2 - z_3| =2$ is satisfied for all the proposed 
forms of CSD, perhaps because one selected the simplest integer valued alignments. 
However, it should be noted that other more complicated but equally 
successful examples were found that violated 
the condition $|z_2 - z_3| =2$ and these were also tabulated
in \cite{King:2013iva}.

\end{document}